\documentclass[10pt]{article}
\usepackage{graphicx}
\usepackage{amsmath}
\usepackage{amssymb}
\usepackage{caption2}
\begin{document}
\bibliographystyle{prsty}
\begin{center}
{\large {\bf \sc{  In-medium  mass modifications of the $D_0$ and $B_0$ mesons  with   the   QCD sum rules  }}} \\[2mm]
Zhi-Gang Wang$^{1}$ \footnote{E-mail,wangzgyiti@yahoo.com.cn.  }, Tao Huang$^{2}$ \footnote{Email: huangtao@ihep.ac.cn}     \\
$^{1}$ Department of Physics, North China Electric Power University, Baoding 071003, P. R. China \\
$^{2}$ Institute of High Energy Physics and Theoretical Physics
Center
for Science Facilities, Chinese Academy of Sciences, Beijing 100049, P.R. China \\
\end{center}

\begin{abstract}
In this article, we calculate the in-medium  mass modifications  of
the scalar  mesons $D_0$ and $B_0$ using the QCD sum rules.  In
calculations, we observe that the  $D_0N$ and $B_0N$ scattering
lengths are about $1.1\,\rm{fm}$ and  $4.1\,\rm{fm}$ respectively,
 the mass-shifts  $\delta M_{D_0}=69\,\rm{MeV}$ and  $\delta
M_{B_0}=217\,\rm{MeV}$, and the $D_0N$ and $B_0N$ interactions are
repulsive.  The  positive   mass-shifts indicate that   the decays
of the higher charmonium states into
  the $D_0\bar{D}_0$ pair are  suppressed.
\end{abstract}


\section{Introduction}

The   in-medium properties of hadrons play an important role  in
understanding the strong interactions, the heavy ion collisions and
the nuclear astrophysics. There have been many experiments devoted
to study the in-medium  hadron properties. The upcoming  FAIR
(Facility for Antiproton and Ion Research) project at GSI
 provides the opportunity to extend the experimental
studies into the charm sector. The CBM (compressed baryonic matter)
collaboration  intends to study the in-medium properties of the
 hadrons, including charmed mesons  \cite{CBM},
 while the $\rm{\bar{P}ANDA}$ collaboration will focus on  the charm spectroscopy,
 and mass and width modifications  of the charmed hadrons
 in the nuclear matter \cite{PANDA}.
  Therefore, the in-medium properties of the charmed mesons
become excellent subjects in  recent years.

The suppression of the $J/\psi$ production in  relativistic  heavy
ion collisions (such as  RHIC or LHC) may be one of the important
signatures to identify the possible phase  transition  to the
quark-gluon plasma \cite{Matsui86}. Although the dissociation of the
$J/\psi$ in the quark-gluon plasma due to color screening reduces
its production,  the effects of hadron absorptions \cite{Hadron AB}
and co-mover interactions \cite{Co-Move} also play an important
role, and one should be careful to
   make  definitive conclusion.
The in-medium modifications of the hadron properties
   can affect  the productions  of the open-charmed mesons and  the $J/\psi$ in
the heavy ion collisions, as the higher charmonium states are
considered as the major source of the $J/\psi$ \cite{Jpsi-Source}.
If the  mass reductions
   of the $D$, $\bar{D}$ mesons are large enough, the
excited charmonium states can decay to the $D\bar{D}$ pair instead
of decaying to the lowest  state $J/\psi$, the $J/\psi$ itself can
also decay to the $D\bar{D}$ pair  at even higher nuclear density
\cite{H-to-Jspi}, although the excited charmonium states
 undergo mass modifications  in the nuclear matter.
The $D$-meson can also be  produced through the reaction   $\psi+N
\rightarrow D+\bar{D}$ in  the heavy ion collisions,
 on the other hand, the $D$-mesons
  couple strongly through the inelastic channels such as
$DN$ $\rightarrow \Lambda_c,\,\Sigma_c$ \cite{Hayashigaki}.

The in-medium mass modifications of the $D,\bar{D}$ mesons have been
studied with  the QCD sum rules \cite{Hayashigaki,DB-SR}, the
quark-meson coupling  model \cite{QMC}, the coupled-channel approach
\cite{C-channel}, etc. The works on the scalar  $D_0$, $\bar{D}_0$
mesons are few \cite{SumRule-D0,C-channel-D0},   it is interesting to study their
mass modifications in the nuclear matter to see whether or not the
decaying of the higher excited charmonium states to the
$D_0\bar{D}_0$ pair is facilitated. Just like the $D$ meson, the
$D_0$ meson
 contains a  charm-quark and a light quark.
The existence of a light quark in the charmed mesons results in much
difference between the in-medium mass modifications of
 the charmed-mesons and charmonium states.
The former have  large contributions  from the light-quark
condensates,  while the latter
  are dominated by the gluon condensates \cite{Hayashigaki,DB-SR,SumRule-D0,Jpsi-etac}.
  In this article, we study the mass modifications of the $D_0,\bar{D}_0$ mesons in the nuclear
matter using the QCD sum rules \cite{SVZ79}, furthermore, we study
the corresponding mass modifications of the $B_0,\bar{B}_0$ mesons
considering the heavy quark symmetry. In Ref.\cite{SumRule-D0}, Hilger and  Kampfer study the in-medium mass
modifications of the scalar  $D_0,\bar{D}_0$ mesons and
 quantify the  $D_0-\bar{D}_0$ mass splitting using the QCD sum rules, the hadronic spectral
 density they choose  differs  from that of the present work.
 In Ref.\cite{C-channel-D0}, Tolos et al study the in-medium properties of the scalar mesons
 $D_0$, $D_{s0}(2317)$ and $X(3700)$ in the coupled-channel approach and reproduce them as dynamically-generated resonances.

The article is arranged as follows:  we study the in-medium mass
modifications of the scalar mesons $D_0$ and $B_0$
 with  the  QCD sum rules in Sec.2; in Sec.3, we present the
numerical results and discussions; and Sec.4 is reserved for our
conclusions.

\section{In-medium mass modifications of the $D_0$ and $B_0$  with  QCD sum rules}

We study the mass modification of the $D_0$ meson in nuclear matter
with the two-point correlation function $\Pi(q)$. In the Fermi gas
approximation for the nuclear matter, the $\Pi(q)$ can be  divided
into  the vacuum part $\Pi_0(q)$  and the static one-nucleon part
$\Pi_N(q)$, which
 is expected to be valid at relatively low nuclear density, and the
$\Pi_{N}(q)$ can be approximated  in the linear density of the
nuclear matter \cite{Hayashigaki,Drukarev1991},
\begin{eqnarray}
\Pi(q) &=& i\int d^{4}x\ e^{iq \cdot x} \langle
T\left\{J(x)J^{\dag}(0)\right\} \rangle_{\rho_N} = \Pi_{0}(q)
+\Pi_{N}(q)\simeq\Pi_{0}(q)+ \frac{\rho_N}{2M_N}T_{N}(q)\, ,
 \end{eqnarray}
where the $\rho_N$ denotes the density of the nuclear matter, and
the forward scattering amplitude $T_{N}(q)$ is defined as
\begin{eqnarray}
T_{N}(\omega,\mbox{\boldmath $q$}\,) &=&i\int d^{4}x e^{iq\cdot x}\langle N(p)|
T\left\{J(x)J^{\dag}(0)\right\} |N(p) \rangle\, ,
\end{eqnarray}
the $J(x)$ denotes the isospin averaged current,
\begin{eqnarray}
 J(x) &=&J^\dag(x) =\frac{\bar{c}(x) q(x)+\bar{q}(x) c(x)}{2},
\end{eqnarray}
 the $q$ denotes the $u$ or $d$ quark, the $q^{\mu}=(\omega,\mbox{\boldmath $q$}\,)$ is the four-momentum carried by
the  current $J(x)$, the $|N(p)\rangle$ denotes the isospin and
spin averaged static nucleon state with the four-momentum $p =
(M_N,0)$, and  $\langle N(\mbox{\boldmath $p$})|N(\mbox{\boldmath
$p$}')\rangle = (2\pi)^{3} 2p_{0}\delta^{3}(\mbox{\boldmath
$p$}-\mbox{\boldmath $p$}')$ \cite{Hayashigaki}. The terms
proportional to $p_F^4$, $p_F^5$, $p_F^6$, $\cdots$ can be neglected
at the normal nuclear matter with the saturation density
$\rho_N=\rho_0=\frac{2p_F^3}{3\pi^2}$, as the Fermi momentum
$p_F=0.27\,\rm{GeV}$ is a small quantity \cite{Drukarev1991}.

In the limit $\mbox{\boldmath $q$}\rightarrow {\bf 0}$, the
$T_{N}(\omega,\mbox{\boldmath $q$}\,)$ can be related to the $D_0N$
scattering $T$-matrix, ${{\cal T}_{D_0N}}(M_{D_0},0) =
8\pi(M_N+M_{D_0})a_{D_0}$, where the  $a_{D_0}$ is the $D_0N$
scattering length. Near the pole position of the $D_0$-meson, the
phenomenological spectral density $\rho(\omega,0)$ can be
parameterized with three unknown   parameters $a,b$ and $c$
\cite{Hayashigaki},
\begin{eqnarray}
\rho(\omega,0) &=& -\frac{f_{D_0}^2M_{D_0}^2}{\pi} \mbox{Im} \left[\frac{{{\cal T}_{D_0N}}(\omega,{\bf 0})}{(\omega^{2}-
M_{D_0}^2+i\varepsilon)^{2}} \right]+ \cdots \,, \\
&=& a\,\frac{d}{d\omega^2}\delta(\omega^{2}-M_{D_0}^2) +
b\,\delta(\omega^{2}-M_{D_0}^2) + c\,\delta(\omega^{2}-s_{0})\, ,
\end{eqnarray}
where the  decay constant $f_{D_0}$ is defined by   $\langle
0|J(0)|D_0(k)+\bar{D}_0(k)\rangle = f_{D_0}M_{D_0}$, the terms
denoted by $\cdots$ represent the continuum contributions. The first
term denotes the double-pole term, and corresponds  to the on-shell
effect of the $T$-matrix, $a=-8\pi(M_N+M_{D_0})a_{D_0}
f_{D_0}^2M_{D_0}^2$; the second term denotes  the single-pole term,
and corresponds to the off-shell effect of the $T$-matrix; and the
third term denotes   the continuum term or the  remaining effects,
where the $s_{0}$ is the continuum threshold. Then the mass-shift of
the $D_0$-meson can be approximated as
\begin{eqnarray}
\delta M_{D_0} &=&2\pi\frac{M_{N}+M_{D_0}}{M_NM_{D_0}}\rho_N a_{D_0}\, .
\end{eqnarray}

In the low energy limit $\omega\rightarrow 0$, the
$T_{N}(\omega,{\bf 0})$ is equivalent to the Born term $T_{N}^{\rm
Born}(\omega,{\bf 0})$, i.e. $T_{N}(0)=T_{N}^{\rm Born}(0)$. We take
into account the Born term at the phenomenological side,
\begin{eqnarray}
T_{N}(\omega^2)&=&T_{N}^{\rm
Born}(\omega^2)+\frac{a}{(M_{D_0}^2-\omega^2)^2}+\frac{b}{M_{D_0}^2-\omega^2}+\frac{c}{s_0-\omega^2}
\, ,
\end{eqnarray}
with the constraint
\begin{eqnarray}
\frac{a}{M_{D_0}^4}+\frac{b}{M_{D_0}^2}+\frac{c}{s_0}&=&0 \, .
\end{eqnarray}

The contributions from the intermediate spin-$\frac{3}{2}$ charmed
baryon states are zero in the soft-limit $q_\mu \to 0$
\cite{Wangzg}, and we only take into account the intermediate
spin-$\frac{1}{2}$ charmed  baryon states in calculating the Born term.
The isospin states of the $D_0$-mesons have the
  $D_0^0N$ and $D_0^+N$ interactions,
\begin{eqnarray}
D_0^0(c\bar{u})+p(uud)\ \mbox{or}\ n(udd)&\longrightarrow&
\Lambda_c^+,\Sigma_c^+(cud)\ \mbox{or}\ \Sigma_c^0(cdd) \, ,\nonumber\\
 D_0^+(c\bar{d})+p(uud)\ \mbox{or}\ n(udd)
&\longrightarrow& \Sigma_c^{++}(cuu)\ \mbox{or}\
\Lambda_c^+,\Sigma_c^+(cud)\, ,
\end{eqnarray}
where   $M_{\Lambda_c}=2.286\,\rm{GeV}$   and
$M_{\Sigma_c}=2.454\,\rm{GeV}$ \cite{PDG}. We can take $M_H\approx
2.4\,\rm{GeV}$ as the average value, where the $H$ means either
$\Lambda_c^+$, $\Sigma_c^+$, $\Sigma_c^{++}$ or $\Sigma_c^0$. It is
straightforward to calculate the Born term by writing down the
Feynman  diagram, the result is
\begin{eqnarray}
T_{N}^{\rm
Born}(\omega,{\bf0})&=&\frac{2f_{D_0}^2M_{D_0}^2M_N(M_H-M_N)g_{D_0NH}^2}
{\left[\omega^2-(M_H-M_N)^2\right]\left[\omega^2-M_{D_0}^2\right]^2}\,,
\end{eqnarray}
where the $g_{D_0NH}$ denotes the  strong coupling constants
$g_{D_0N\Lambda_c}$
 and  $g_{D_0N\Sigma_c}$.
On the other hand, there are no inelastic channels for the
$\bar{D}_0^0  N$ and $D_0^- N$ interactions, i.e. $T_{N}^{\rm
Born}(0)=0$.

 We carry out the operator product expansion to the  condensates  up to dimension-5
 at the large space-like  region in the nuclear matter.
Once analytical results at the level of quark-gluon degree's of  freedom are obtained,
 then we set  $\omega^2=q^2$, and take the
quark-hadron duality and perform the Borel transform  with respect
to the variable $Q^2=-\omega^2$, finally   obtain  the following sum
rule:
\begin{eqnarray}
&& a \left\{\frac{1}{M^2}e^{-\frac{M_{D_0}^2}{M^2}}-\frac{s_0}{M_{D_0}^4}e^{-\frac{s_0}{M^2}}\right\}
+b \left\{e^{-\frac{M_{D_0}^2}{M^2}}-\frac{s_0}{M_{D_0}^2}e^{-\frac{s_0}{M^2}}\right\}
+ \frac{2f_{D_0}^2M_{D_0}^2M_N(M_H-M_N)g_{D_0NH}^2}{(M_H-M_N)^2-M_{D_0}^2}\nonumber\\
&&\left\{ \left[\frac{1}{(M_H-M_N)^2-M_{D_0}^2}-\frac{1}{M^2}\right]
e^{-\frac{M_{D_0}^2}{M^2}}-\frac{1}{(M_H-M_N)^2-M_{D_0}^2}e^{-\frac{(M_H-M_N)^2}{M^2}}\right\}=\left\{\frac{m_c\langle\bar{q}q\rangle_N}{2}\right.\nonumber\\
&&\left.-\langle q^\dag i D_0q\rangle_N+\frac{m_c^2\langle q^\dag i D_0q\rangle_N}{M^2}-\frac{2m_c\langle \bar{q} i D_0 i D_0q\rangle_N}{M^2}+\frac{m_c^3\langle \bar{q} i D_0 i D_0q\rangle_N}{M^4}\right\}e^{-\frac{m_c^2}{M^2}}\nonumber\\
&&+\frac{1}{16}\langle\frac{\alpha_sGG}{\pi}\rangle_N\int_0^1dx \left(1+\frac{\widetilde{m}_c^2}{M^2}\right)e^{-\frac{\widetilde{m}_c^2}{M^2}}
-\frac{1}{48M^4}\langle\frac{\alpha_sGG}{\pi}\rangle_N\int_0^1\frac{1-x}{x}\widetilde{m}_c^4e^{-\frac{\widetilde{m}_c^2}{M^2}}\, ,
\end{eqnarray}
where $\widetilde{m}_c^2=\frac{m_c^2}{x}$.

Differentiate  above equation with respect to  $\frac{1}{M^2}$, then
eliminate the
 parameter $b$, we can obtain the sum rule for
 the parameter $a$. With the simple replacements $m_c \to m_b$, $D_0 \to B_0$, $\Lambda_c \to \Lambda_b$ and $\Sigma_c \to \Sigma_b$,
 we can obtain the corresponding sum rule for
 the mass modification of the $B_0$ meson in the nuclear matter.

\section{Numerical results and discussions}
In calculations, we have assumed that  the linear density
approximation   is valid at the low nuclear  density, for a general
condensate in the nuclear matter,
$\langle{\cal{O}}\rangle_{\rho_N}=\langle0|{\cal{O}}|0\rangle+\frac{\rho_N}{2M_N}\langle
N|{\cal{O}}|N\rangle=\langle{\cal{O}}\rangle_0+\frac{\rho_N}{2M_N}\langle
{\cal{O}}\rangle_N$. The input parameters are taken as
  $\langle\bar{q} q\rangle_N={\sigma_N \over m_u+m_d } (2M_N)$,
 $\langle\frac{\alpha_sGG}{\pi}\rangle_N= - 0.65 \,{\rm {GeV}} (2M_N)$,
$\langle q^\dagger iD_0 q\rangle_N=0.18 \,{\rm{GeV}}(2M_N)$,
$\langle\bar{q}g_s\sigma G q\rangle=3.0\,{\rm GeV}^2(2M_N) $,
$\langle \bar{q} iD_0iD_0
q\rangle_N+{1\over8}\langle\bar{q}g_s\sigma G
q\rangle_N=0.3\,{\rm{GeV}}^2(2M_N)$, $m_u+m_d=12\,\rm{MeV}$,
$\sigma_N=45\,\rm{MeV}$, $M_N=0.94\,\rm{GeV}$, and
$\rho_N=(0.11\,\rm{GeV})^3$ \cite{C-parameter}.

The hadronic parameters $M_{D_0}$, $M_{B_0}$, $f_{D_0}$, $f_{B_0}$
are determined by the conventional two-point correlation functions
$\Pi_0(q)$ using the QCD sum rules,  $M_{D_0}=2.355\,\rm{GeV}$,
$M_{B_0}=5.74\,\rm{}GeV$, $f_{D_0}=0.334\,\rm{GeV}$,
$f_{B_0}=0.28\,\rm{GeV}$ with the threshold parameters
$s^0_D=8.0\,\rm{GeV}^2$ and $s^0_B=39.0\,\rm{GeV}^2$, respectively.
For the observed scalar meson $D_0(2400)$, the value
$M_{D_0}=2.355\,\rm{GeV}$ is consistent with the average  of the
experimental data $M_{D_0^0}=2318\,\rm{MeV}$ and
$M_{D_0^{\pm}}=2403\,\rm{MeV}$ \cite{PDG}. The uncertainties come
from the
 hadronic parameters  $f_{D_0}$ and $f_{B_0}$ can
be approximated as  $\frac{2\delta f_{D_0}} {f_{D_0}}$ and
$\frac{2\delta f_{B_0}} {f_{B_0}}$, respectively.

In Fig.1, we plot the mass-shifts  $\delta M$ versus the Borel
parameter $M^2$. From the figure, we can see that the values of the
mass-shifts  are rather stable with variations  of the Borel
parameter at the intervals $M^2=(6.1-7.4)\,\rm{GeV}^2$ and
$(33-39)\,\rm{GeV}^2$ in the charmed and bottom channels,
respectively, the uncertainties originate from the Borel parameter
$M^2$ are less than $1\%$.
  Furthermore,  the  exponential  factor $e^{-\frac{s_0}{M^2}}<e^{-1}$ at those intervals, the continuum contributions are greatly suppressed. The main comtribtuions come from the terms $m_c\langle\bar{q}q\rangle_N$ and   $m_b\langle\bar{q}q\rangle_N$, the operator product expansion is certainly convergent.

In the Borel windows, the mass-shifts $\delta M_{D_0}=61\,\rm{MeV}$, $65\,\rm{MeV}$, $68\,\rm{MeV}$, $72\,\rm{MeV}$, $75\,\rm{MeV}$, $79\,\rm{MeV}$
and $\delta M_{B_0}=213\,\rm{MeV}$, $215\,\rm{MeV}$, $217\,\rm{MeV}$, $219\,\rm{MeV}$, $221\,\rm{MeV}$, $223\,\rm{MeV}$ respectively  at the values
$g^2=0$, $20$, $40$, $60$, $80$, $100$, where the $g$ denotes  the strong coupling constants $g_{D_0N\Lambda_c}$, $g_{D_0N\Sigma_c}$,
$ g_{B_0N\Lambda_b}$, $g_{B_0N\Sigma_b}$, the mass-shifts increase monotonously with increase of the squared strong coupling constants
$g^2$.

The calculations based on the QCD sum rules indicate that the values of the strong coupling constants
 $g_{NN\sigma(600)}=12\pm2$ and $g_{NNa_0(980)}=11\pm2$ \cite{TMAliev},
 which are of the same magnitude of the phenomenological value of the
strong coupling constant $g_{NN\pi}=13.5$. On the other hand, the
value of the strong coupling constant $g_{DN\Lambda_c}=6.74$ from
the QCD sum rules  is much smaller \cite{Nielsen98}. In this
article, we take the approximation $g_{D_0N\Lambda_c}\approx
g_{D_0N\Sigma_c}\approx g_{B_0N\Lambda_b}\approx
g_{B_0N\Sigma_b}\approx6.74$, and obtain the values $\delta
M_{D_0}=69\,\rm{MeV}$,  $\delta M_{B_0}=217\,\rm{MeV}$,
$a_{D_0}=1.1\,\rm{fm}$ and $a_{B_0}=4.1\,\rm{fm}$.

The positive scattering lengths indicate that the $D_0N$ and $B_0N$
interactions are repulsive, it is difficult  to form the $D_0N$ and
$B_0N$ bound states. Which are in contrast to the $DN$ and $BN$
interactions, where the negative scattering lengths indicate   that
the $DN$ and $BN$ interactions are attractive, it is possible to
form the $DN$ and $BN$ bound states.

Due to positive  mass-shift $\delta M_{D_0}$, the decays of the high
charmonium states to the $D_0\bar{D}_0$ pair obtain additional
suppression in the phase space, and  the decay into the lowest state
$J/\psi$ is preferred. While the negative
  mass-shift  $\delta M_{D}$ indicate that the high charmonium states decays to
  the $D\bar{D}$ pair are facilitated, and the production of the
 $J/\psi$ is suppressed. The $J/\psi$ production does not obtain additional suppression due to
 mass modification of the scalar meson $D_0(2400)$
  in the  nuclear matter.

\begin{figure}
 \centering
 \includegraphics[totalheight=6cm,width=7cm]{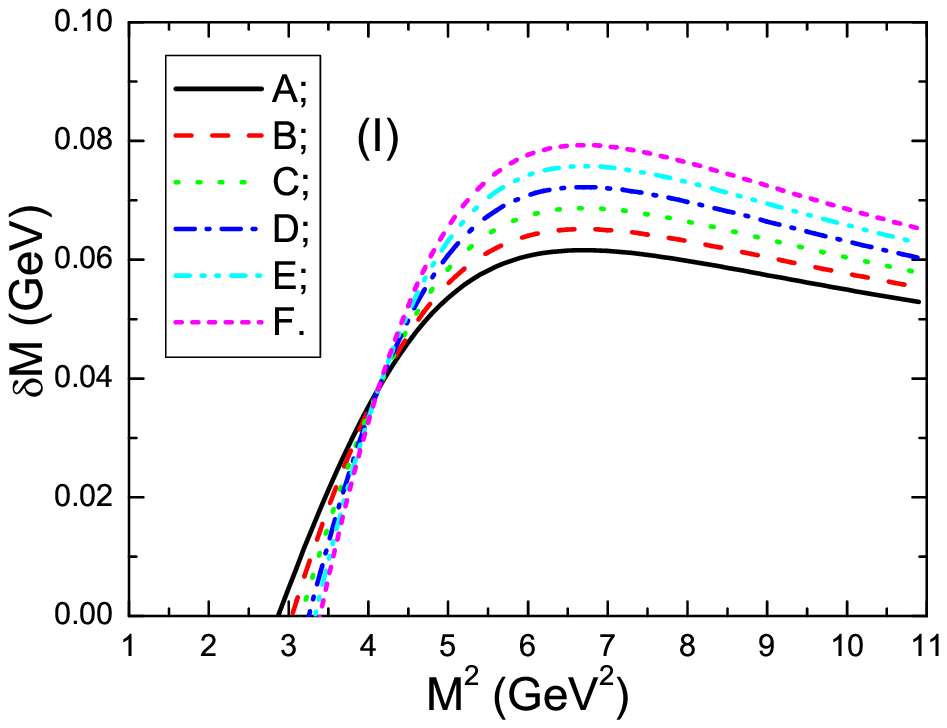}
 \includegraphics[totalheight=6cm,width=7cm]{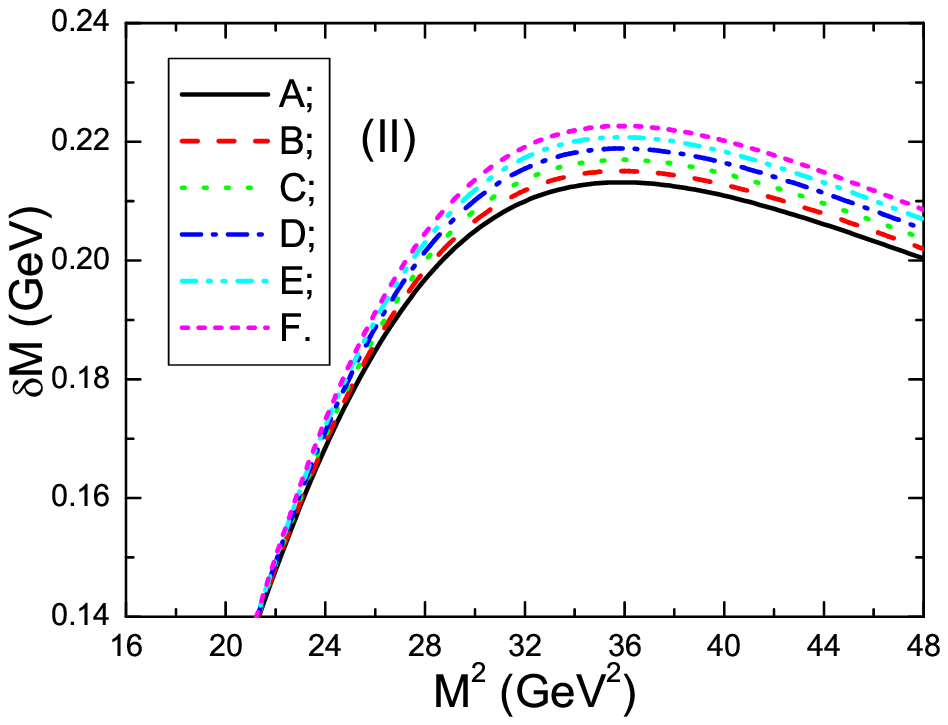}
 \caption{The   mass-shifts $\delta M$ versus the Borel parameter $M^2$, the (I) and (II) denote the $D_0$ and $B_0$ mesons respectively,
 the $A$, $B$, $C$, $D$, $E$ and $F$ correspond to the values  $g^2=0$, $20$, $40$, $60$, $80$ and $100$, respectively. }
\end{figure}

In the present work and Ref.\cite{Hayashigaki},  the correlation functions  $\Pi(q)$ are  divided
into  the vacuum part   and the static one-nucleon part, and  the nuclear matter  induced effects are extracted  explicitly,
  while in Refs.\cite{DB-SR,SumRule-D0},   the pole terms of the     hadronic spectral densities  are  parameterized
as $\frac{\rm{Im}\Pi(\omega,0)}{\pi}=F_{+}\delta(\omega-M_{+})-F_{-}\delta(\omega+M_{-})$, where $M_{\pm}=M\pm\Delta M$ and $F_{\pm}=F\pm\Delta F$,
and  the sum rules for the   mass center $M$ and the mass splitting $\Delta M$ are obtained. For the pseudoscalar $D,\bar{D}$ mesons,
 Hayashigaki  obtains the mass-shift $\delta M_{D}=-50\,\rm{MeV}$ \cite{Hayashigaki},
while  Hilger, Thomas and Kampfer obtain the mass-shift
$\delta M_{D}=+45\,\rm{MeV}$ \cite{DB-SR}.
For scalar $D_0,\bar{D}_0$  mesons, the  mass-shift  $\delta M_{D_0}=M-M_{D_0}<0$ obtained by
Hilger and Kampfer \cite{SumRule-D0} differs  from the result   in the present work.

In those studies, irrespective of the parameterizations of the hadronic spectral densities,
derivatives  with respect to $1/M^2$ are used to obtain additional QCD sum rules so as to take special superpositions to delete the unknown parameters and
extract the explicit expressions of the mass-shifts. The QCD sum rules from the derivatives are not necessarily work well, as the QCD sum rules are
just a QCD model, the predictions  can vary with the special assumptions. Furthermore, the isospin currents and isospin-averaged currents can also lead to
 differences in the hadronic spectral densities. All those predictions can be confronted with the experimental  data in the future.

In Ref.\cite{C-channel-D0}, Tolos et al study the in-medium properties of the scalar mesons
 $D_0(2400)$, $D_{s0}(2317)$ and $X(3700)$ in the coupled-channel approach and reproduce  them as dynamically-generated resonances,
 and observe that the $D_{s0}(2317)$ and $X(3700)$   enlarge their widths to   the order of 100 and 200 $\rm{MeV}$ respectively at the normal
nuclear matter, and
the $D_0(2400)$ meson obtains  an extra widening from the already large width, due to the $D$ meson absorptions
in the nuclear matter via the $DN$ and $DNN$ inelastic reactions.  In the vacuum, the mass and width of the $D_0(2400)$ are
 $M_{D_0^0}=(2318\pm29)\,\rm{MeV}$, $M_{D_0^\pm}=(2403\pm14\pm35)\,\rm{MeV}$,   $\Gamma_{D_0^0}=(267\pm40)\,\rm{MeV}$,
 and $\Gamma_{D_0^\pm}=(283\pm24\pm34)\,\rm{MeV}$, respectively \cite{PDG}. For $D_0^0(2400)$, the spin-parity $J^P=0^+$ assignment is favored,
 while the spin-parity $J^P$ of the $D_0^{\pm}(2400)$ still needs  confirmation \cite{PDG}.   The mass-shifts $\delta
M_{D_0}=69\,\rm{MeV}$ and $\delta M_{D}=-50\,\rm{MeV}$ obtained in the present work and Ref.\cite{Hayashigaki} respectively
favor the decays $D_0 \to D \pi$, we can expect that the in-medium width of the $D_0(2400)$ is larger, and our prediction is compatible
with the observation of Ref.\cite{C-channel-D0}.
  However, the large width  disfavors the experimental observation of the relative small mass-shift.

\section{Conclusion}
In this article, we calculate the in-medium mass-shifts of the
scalar mesons $D_0$ and $B_0$ using the QCD sum rules. At the low
density of the nuclear matter, we can take the linear approximation,
and extract  the mass-shifts explicitly. Our numerical results
indicate that the  $D_0N$ and $B_0N$ scattering lengths are
$a_{D_0}=1.1\,\rm{fm}$ and $a_{B_0}=4.1\,\rm{fm}$, respectively, the
$D_0N$ and $B_0N$ interactions are repulsive,
  the mass-shifts are  $\delta M_{D_0}=69\,\rm{MeV}$ and  $\delta M_{B_0}=217\,\rm{MeV}$.
   The  positive  mass-shifts indicate that the $J/\psi$ production does not  obtain
    additional suppression due to the mass modification of the scalar
    meson $D_0(2400)$  in the  nuclear matter.

\section*{Acknowledgments}
This  work is supported by National Natural Science Foundation,
Grant Number 11075053,  and the Fundamental
Research Funds for the Central Universities.


\begin{thebibliography}{99}

\bibitem{CBM} B. Friman et al, "The CBM Physics Book: Compressed Baryonic Matter in Laboratory
Experiments", Springer Heidelberg.

\bibitem{PANDA} M. F. M. Lutz et al, arXiv:0903.3905.


\bibitem{Matsui86} T. Matsui and H. Satz,  Phys. Lett. {\bf B178} (1986) 416;
R. Vogt, Phys. Rept. {\bf 310} (1999) 197.


\bibitem{Hadron AB}  B. Zhang et al, Phys. Rev. {\bf C62} (2000) 054905;
W. Cassing and E. L. Bratkovskaya, Nucl. Phys. {\bf A623} (1997)
570; E. L. Bratkovskaya, W. Cassing and H. Stoecker, Phys. Rev. {\bf
C67} (2003) 054905.


\bibitem{Co-Move}   A. Capella and E. G. Ferreiro, Eur. Phys. J. {\bf C42} (2005) 419;
 N. Armesto et al,  Nucl. Phys. {\bf A698} (2002) 583;
 S. Gavin and  R. Vogt, Phys. Rev. Lett. {\bf 78} (1997) 1006.


\bibitem{Jpsi-Source} L. Antoniazzi et al, Phys. Rev. Lett. {\bf 70} (1993) 383;
Y. Lemoigne et al,  Phys. Lett. {\bf B113} (1982) 509.

\bibitem{H-to-Jspi}  Y. S. Golubeva et al,  Eur. Phys. J. {\bf A17} (2003) 275.


\bibitem{Hayashigaki}   A. Hayashigaki, Phys. Lett. {\bf B487} (2000) 96.

\bibitem{DB-SR} T. Hilger, R. Thomas and B. Kampfer, Phys. Rev. {\bf C79} (2009) 025202.

\bibitem{QMC} K. Tsushima  et al,  Phys. Rev. {\bf C59} (1999) 2824;
  A. Sibirtsev, K. Tsushima and A. W. Thomas, Eur. Phys. J. {\bf A6} (1999)  351.

\bibitem{C-channel}  L. Tolos, J. Schaffner-Bielich and A. Mishra, Phys. Rev. {\bf C70} (2004) 025203;
 L. Tolos, J. Schaffner-Bielich and H. Stoecker, Phys. Lett. {\bf B635} (2006)
 85; T. Mizutani and A. Ramos, Phys. Rev. {\bf C74} (2006) 065201;  L. Tolos, A. Ramos and T.  Mizutani,
 Phys. Rev. {\bf C77} (2008) 015207;  L. Tolos,  C. Garcia-Recio and J. Nieves,  Phys. Rev. {\bf C80} (2009)  065202.

\bibitem{SumRule-D0}  T. Hilger and B. Kampfer, Nucl. Phys. Proc. Suppl. {\bf 207-208} (2010) 277.

\bibitem{C-channel-D0} L. Tolos, R. Molina, D. Gamermann and E. Oset, Nucl. Phys. {\bf A827} (2009) 249c;
 L. Tolos, D. Gamermann, R. Molina, E. Oset and A. Ramos, arXiv:0905.1850.




\bibitem{Jpsi-etac} A. Hayashigaki,  Prog. Theor. Phys. {\bf 101} (1999) 923;
F. Klingl et al, Phys. Rev. Lett. {\bf 82} (1999) 3396; S. Kim and
S. H. Lee, Nucl. Phys. {\bf A679} (2001) 517; C. M. Ko and S. H.
Lee, Phys. Rev. {\bf C67} (2003) 038202; A. Kumar and A. Mishra,
Phys. Rev. {\bf C82} (2010) 045207.


\bibitem{SVZ79} M. A. Shifman, A. I.  Vainshtein and V. I. Zakharov,
Nucl. Phys. {\bf B147} (1979) 385; L. J. Reinders, H. Rubinstein and
S. Yazaki, Phys. Rept. {\bf 127} (1985) 1.

\bibitem{Drukarev1991} E. G. Drukarev and E. M. Levin,   Prog. Part. Nucl. Phys. {\bf 27} (1991)
77; E. G. Drukarev, M. G. Ryskin and V. A. Sadovnikova, Prog. Part.
Nucl. Phys. {\bf 47} (2001) 73.


\bibitem{Wangzg} Z. G. Wang,  Eur. Phys. J. {\bf C57} (2008) 711;
Z. G. Wang,  Eur. Phys. J. {\bf C61} (2009) 299.



\bibitem{PDG}  K. Nakamura et al, J. Phys. {\bf G37} (2010)  075021.

\bibitem{C-parameter} T. D. Cohen et al, Prog. Part. Nucl. Phys. {\bf 35} (1995)
221; X. M. Jin et al,  Phys. Rev. {\bf C47} (1993) 2882; X. M. Jin
et al,  Phys. Rev. {\bf C49} (1994) 464.

\bibitem{TMAliev} T. M. Aliev and M. Savci, Phys. Rev. {\bf D75} (2007) 045006.

\bibitem{Nielsen98} F. S. Navarra and  M. Nielsen, Phys. Lett. {\bf B443} (1998) 285.

\end{thebibliography}
\end{document}